\documentclass[superscriptaddress,prl]{revtex4}
\RequirePackage{graphicx}

\begin{document}
\title{Solitary waves in plasmonic Bragg gratings}

\author{Ildar\,R.~Gabitov}
\email{gabitov@math.arizona.edu} \affiliation{Department of
Mathematics, University of Arizona\\617 North Santa Rita Avenue,
Tucson, AZ 85721, USA} \affiliation{L.D. Landau Institute for
Theoretical Physics, Russian Academy of Sciences\\2 Kosygin Street,
Moscow, 119334, Russian Federation}

\author{Alexander\,O.~Korotkevich}
\email{kao@itp.ac.ru} \affiliation{L.D. Landau Institute for
Theoretical Physics, Russian Academy of Sciences\\2 Kosygin Street,
Moscow, 119334, Russian Federation}

\author{Andrei\,I.~Maimistov}
\email{maimistov@pico.miphi.ru} \affiliation{Department of Solid
State Physics, Moscow Engineering Physics Institute\\Moscow, 115409,
Russian Federation}

\author{Joseph\,B.~McMahon}
\email{jmcmahon@math.arizona.edu} \affiliation{Program in Applied
Mathematics, University of Arizona\\617 North Santa Rita Avenue,
P.O. Box 210089, Tucson, AZ 85721-0089, USA}

\begin{abstract}
Light propagation in a Bragg periodic structure containing thin films with
metallic nanoparticles is studied. Plasmonic resonance frequency, Bragg
frequency, and light carrier frequency are assumed to be close. Exact
solutions describing solitary gap-waves are found, and a light arrest
phenomenon due to nonlinearity of plasmonic oscillations is studied.
\end{abstract}

\maketitle

\affiliation{Department of Mathematics, University of Arizona\\617 North
Santa Rita Avenue, Tucson, AZ 85721, USA}
\affiliation{L.D. Landau Institute for Theoretical Physics, Russian Academy
of Sciences\\2 Kosygin Street,
Moscow, 119334, Russian Federation}

\affiliation{L.D. Landau Institute for Theoretical Physics, Russian Academy of
Sciences\\2 Kosygin Street,
Moscow, 119334, Russian Federation}

\affiliation{Department of Solid State Physics, Moscow Engineering Physics
Institute\\Moscow, 115409, Russian Federation}

\affiliation{Program in Applied Mathematics, University of Arizona\\617 North
Santa Rita Avenue,
P.O. Box 210089, Tucson, AZ 85721-0089, USA}

\section{Introduction}

The field of photonic crystals, driven by its importance for
applications in various areas of photonics, has attracted
considerable attention during last the ten years~\cite{1}-\cite{10}.
The one-dimensional \textit{resonant Bragg grating}
\cite{1}-\cite{5} or \textit{resonantly absorbing Bragg reflector}~
\cite{6}-\cite{8} has been of particular interest. An idealized
model of a resonant Bragg grating often considers a linear
homogeneous dielectric host medium containing an array of thin films
with resonant atoms or molecules. The thickness of each film is much
less than the wavelength of the electromagnetic wave propagating
through such a structure. The dynamics of ultrashort light pulses in
the gratings containing films embedded with two-level atoms of the
grating has been studied in~\cite{1}-\cite{8}. The transition
frequencies of atoms are assumed to be near the Bragg frequency.
Light interaction with such gratings was described by equations of
reduced Maxwell-Bloch type. This work demonstrated self-induced
transparency in these gratings and the existence of the solitary
waves in such structures~ \cite{1,4,6}. It was also shown~\cite{8}
that bright as well as dark solitons can exist in the spectral gap.

Progress in nanofabrication of novel optical materials has allowed
the design of metal-dielectric nanocomposite materials, which have
the ability to sustain nonlinear plasmonic oscillations. An example
of such a material is a dielectric with embedded metallic
nanoparticles~\cite{R97}-\cite{HRF86}. In this paper we consider
ultrashort pulse interaction with Bragg structures containing thin
films with metallic nanoparticles. The Bragg resonance frequency and
carrier wave frequency in our considerations are assumed to be close
to the plasmonic resonance frequency of the nanoparticles. Losses in
realistic plasmonic oscillations are of considerable importance.
Here we consider an idealized case, wherein pulse duration is much
shorter than the characteristic time of losses, so that the effects
of losses can be neglected. In the limit of the slowly-varying
envelope approximation we derive governing equations for two
counter-propagating electromagnetic waves interacting with plasmonic
oscillation-induced medium polarization. This system of equations
represents the two-wave Maxwell-Duffing type model. We find exact
solitary wave solutions of this system and, via computer
simulations, analyze stability of these solutions.

\section{Basic Equations}

We consider a grating consisting of an array of thin films which are
embedded in a linear dielectric medium. In our derivation of the governing
equations we follow~\cite{1}-\cite{8}, wherein Bragg resonance arises if the
distance between successive films is $a = \left(\lambda/2\right)m$, $m = 1,
2, 3, \ldots$.

It was shown in~\cite{Ma06,Ma06_1} that counter-propagating electric field
waves $\mathcal{A}$ and $\mathcal{B}$ in the slowly-varying envelope
approximation satisfy the following system of equations:
\begin{eqnarray}
i\left( \frac{\partial }{\partial x}+\frac{1}{v_{g}}\frac{\partial }{
\partial t}\right) \mathcal{A}-\frac{q_{2}}{2}\frac{\partial ^{2}}{\partial
t^{2}}\mathcal{A}+\Delta q_{0}\mathcal{A} &=&-\frac{2\pi \omega _{0}}{c\sqrt{
\varepsilon }}\langle\mathcal{P}\rangle,  \label{21.1} \\
i\left( \frac{\partial }{\partial x}-\frac{1}{v_{g}}\frac{\partial }{
\partial t}\right) \mathcal{B}+\frac{q_{2}}{2}\frac{\partial ^{2}}{\partial
t^{2}}\mathcal{B}-\Delta q_{0}\mathcal{B} &=&+\frac{2\pi \omega _{0}}{c\sqrt{
\varepsilon }}\langle\mathcal{P}\rangle,  \label{21.2}
\end{eqnarray}
where $\Delta q_{0}=q_{0}-2\pi /a$ is the mismatch between the
carrier wavenumber and the Bragg resonant wavenumber. The
description of the evolution of material polarization in the
slowly-varying amplitude approximation requires modeling of the thin
films' response to an external light field. The dielectric
properties can be attributed to plasmonic oscillations, which are
modeled by Lorentz oscillators. The simplest generalizations of this
model include anharmonicity of plasmonic
oscillations~\cite{R16,LGMS}. In this paper we consider an array of
thin films containing metallic nanoparticles which have cubic
nonlinear response to external fields~\cite{R97,DBNS04,R16}.

The macroscopic polarization $P$ is governed by the equation
\[
\frac{\partial ^{2}P}{\partial t^{2}}+\omega _{d}^{2}P+\Gamma
_{a}\frac{\partial P}{\partial t}+\kappa P^{3}=\frac{\omega
_{p}^{2}}{4\pi }E,
\]
where $\omega _{p}$ is plasma frequency and $\omega _{d}$ is the
transition frequency between energy levels resulting from the
dimensional quantization.   Losses of the plasmonic oscillations are
taken into account by the parameter $\Gamma _{a}$. It is assumed
that the duration of the electromagnetic pulse is small enough that
dissipation effects can be neglected.

Starting from the slowly-varying envelope approximation, standard
manipulation leads to
\begin{equation}
i\frac{\partial \mathcal{P}}{\partial t}+(\omega _{d}-\omega_{0})\mathcal{P}
+ \frac{3\kappa }{2\omega_{0}}|\mathcal{P}|^{2}\mathcal{P} = -\frac{\omega
_{p}^{2}}{8\pi\omega _{0}}\mathcal{E}_{int}(x,t).  \label{anhar3}
\end{equation}
Terms varying rapidly in time, which are proportional to $\exp (\pm 3i\omega
_{0}t)$, are neglected. In this equation $\mathcal{E}_{int}$ is the electric
field interacting with metallic nanoparticles. In the problem under
consideration we have $\mathcal{E}_{int}=\mathcal{A}+\mathcal{B}$ .

Due to the limitations of nanofabrication, the sizes and shapes of
nanoparticles are not uniform. In practice, deviation from a
perfectly spherical shape has a much larger impact on a
nanoparticle's resonance frequency than does variation in diameter.
This shape variation causes a broadening of the resonance line. The
broadened spectrum is characterized by a probability density
function $g(\Delta \omega )$ of deviations $\Delta \omega $ from
some mean value $\omega_{res}$. When computing the total
polarization, all resonance frequencies must be taken into account.

The contributions of the various resonance frequencies are weighted
according to the probability density function $g(\Delta \omega )$;
the weighted average is denoted by $\langle \mathcal{P}\rangle $ in
equations~(\ref{21.1}),(\ref{21.2}). In what follows, $n(\omega
_{0})$ denotes the refractive index of the medium containing the
array of thin films, and $ n_{np}$ is the effective density of the
resonant nanoparticles in films.  The effective density is equal to
$n_{np}=N_{np}(\ell _{f}/a)$, where $ N_{np}$ is the bulk density of
nanoparticles, $\ell _{f}$ is the width of a film, and $a$ is the
lattice spacing.

We study a medium-light interaction in which resonance is the dominant
phenomenon. As such, the length of the sample is smaller than the
characteristic dispersion length. In this case the temporal second
derivative terms in equations~(\ref{21.1},\ref{21.2}) can be omitted. The
resulting equations are the two-wave Maxwell-Duffing equations. They can be
rewritten in dimensionless form using the following rescaling:
\[
e_{1} = \mathcal{A}/A_{0},\; e_{2} = \mathcal{B}/A_{0},\; p = (4\pi \omega
_{0}/[\sqrt{\varepsilon }\omega _{p}A_{0}])\mathcal{P},\; \zeta = (\omega
_{p}/2c)x,\; \tau = t/t_{0}.
\]
Here $t_{0}=2\sqrt{\varepsilon }/\omega _{p}$, while $A_{0}$ is a
characteristic amplitude of counter-propagating fields. In dimensionless
form, the two-wave Maxwell-Duffing equations read
\begin{eqnarray}
i\left( \frac{\partial }{\partial \zeta }+\frac{\partial }{\partial \tau }
\right) e_{1}+\delta e_{1} &=&-\langle p\rangle ,  \nonumber \\
i\left( \frac{\partial }{\partial \zeta }-\frac{\partial }{\partial \tau }
\right) e_{2}-\delta e_{2} &=&+\langle p\rangle ,  \label{SVEPEquations} \\
i\frac{\partial p}{\partial \tau }+\Delta p+\mu |p|^{2}p &=&-(e_{1}+e_{2}),
\nonumber
\end{eqnarray}
where $\mu =(3\kappa \sqrt{\varepsilon }/\omega _{0}\omega_{p})(\sqrt{
\varepsilon }\omega _{p}/4\pi \omega_{0})^{2}A_{0}^{2}$ is a dimensionless
coefficient of anharmonicity, $\delta =2\Delta q_{0}(c/\omega _{p})$ is the
dimensionless mismatch coefficient, $\Delta =2\sqrt{\varepsilon}(\omega
_{d}-\omega _{0})/\omega _{p}$ is the dimensionless detuning of a
nanoparticle's resonance frequency from the field's carrier frequency.

In a coordinate system rotating with angular frequency $\delta$,
\[
e_{1}=f_{1}e^{i\delta \tau },\;\indent e_{2}=f_{2}e^{i\delta \tau
},\; \indent p=qe^{i\delta\tau },
\]
equations~(\ref{SVEPEquations}) become
\begin{eqnarray}
i\left( \frac{\partial }{\partial \zeta }+\frac{\partial }{\partial\tau }
\right) f_{1} &=&-\langle q\rangle ,  \nonumber \\
i\left( \frac{\partial }{\partial \zeta }-\frac{\partial }{\partial\tau }
\right) f_{2} &=&+\langle q\rangle ,  \label{RotatingFrame} \\
i\frac{\partial q}{\partial \tau }+(\Delta -\delta )q+\mu |q|^{2}q & =
&-(f_{1}+f_{2}).  \nonumber
\end{eqnarray}
Further simplification of the system~(\ref{RotatingFrame}) can be achieved
by introducing new variables
\[
f_{s}=-(f_{1}+f_{2}),\indent f_{a} = f_{1}-f_{2},
\]
which allow decoupling of one equation from the system of three equations.
In these new variables the polarization $q$ is coupled with only one field
variable. Simple transformations give
\begin{eqnarray}
\frac{\partial ^{2}f_{a}}{\partial \zeta ^{2}}- \frac{\partial ^{2}f_{a}}{%
\partial \tau ^{2}} & = & 2i \frac{\partial }{\partial \zeta }\langle
q\rangle ,  \label{26.3} \\
\frac{\partial ^{2}f_{s}}{\partial \zeta ^{2}}- \frac{\partial ^{2}f_{s}}{%
\partial \tau ^{2}} & = & 2i \frac{\partial }{\partial \tau }\langle
q\rangle,  \label{26.3.1} \\
i\frac{\partial q}{\partial \tau }+(\Delta -\delta )q + \mu |q|^{2}q & = &
f_{s}.  \label{26.3.2}
\end{eqnarray}
As one can see, we have a coupled system of equations for $f_{s}$ and $q$.

\section{Solitary Wave Solutions}

We consider localized solitary wave solutions
of~(\ref{26.3})-(\ref{26.3.2}) in the limit of narrow spectral line
$\Delta \omega_{g}/\Delta \omega_{s} \ll 1$, where $\Delta
\omega_{s}$ and $\Delta \omega_{g}$ are spectral widths of a signal
and spectral line $g(\Delta\omega)$. In this case the spectral line
can be represented as a Dirac $\delta$-function:
$g(\Delta\omega)=\delta(\Delta\omega)$. Equations~(\ref{26.3.1}),
(\ref {26.3.2}) can then be re-written as follows:
\begin{eqnarray}
\frac{\partial ^{2}f_{s}}{\partial \zeta ^{2}}- \frac{\partial ^{2}f_{s}}{%
\partial \tau ^{2}} & = & 2i\frac{\partial q}{\partial \tau } \\
i\frac{\partial q}{\partial \tau }+(\Delta -\delta )q + \mu |q|^{2}q & = &
f_{s}
\end{eqnarray}
Scaling analysis of this system shows that solitary wave solutions can be
represented as
\begin{eqnarray}
f_{s}&=& f_0 F_{\Omega}\left(\eta \right) =
\frac{1}{\sqrt{\mu}}\left( \frac{
2 v^2}{1-v^2}\right )^{3/4}F_{\Omega}\left(\eta \right),  \nonumber \\
q&=& q_0 Q_{\Omega}\left(\eta \right) = \frac{1}{\sqrt{\mu}}\left(
\frac{2 v^2
}{1-v^2}\right )^{1/4}Q_{\Omega}\left(\eta \right),  \label{Scaling} \\
\eta & = & (\zeta - v\tau)\sqrt{\frac{2}{1-v^2}}.  \nonumber
\end{eqnarray}
Here $v$ is velocity of the solitary wave, $\eta$ is a scale-invariant
parameter in a coordinate system moving with the solitary wave, and
functions $F_{\Omega}$, $Q_{\Omega}$ satisfy the following system of
equations:
\begin{eqnarray}
F_{\Omega}^{\prime\prime}& = & -iQ_{\Omega}^{\prime},  \nonumber \\
-iQ_{\Omega}^{\prime}+ \Omega Q_{\Omega} + |Q_{\Omega}|^2 Q_{\Omega} & = &
F_{\Omega} .  \label{ODE}
\end{eqnarray}
The only dimensionless parameter which remains in the system is
\begin{eqnarray}
\Omega = (\Delta - \delta)\sqrt{\frac{1 - v^2}{2 v^2}},
\end{eqnarray}
which characterizes the deviation of carrier frequency from the plasmonic
frequency $\omega_p$ and the Bragg resonance frequency $\omega_{Br}$.

The system of ordinary differential equations~(\ref{ODE}) has integral of
motion for pulse-like solutions decaying as $|\eta |\rightarrow \infty $
\[
|Q_{\Omega }|^{2}-|F_{\Omega }|^{2}=0.
\]
This allows the following parametrization of solutions:
\[
F_{\Omega }(\eta )=R(\eta )e^{i\phi (\eta )},\indent Q_{\Omega }(\eta
)=R(\eta )e^{i\psi (\eta )},
\]
where $R$, $\phi $, and $\psi $ are real-valued functions satisfying
\begin{eqnarray}
R^{\prime } &=&-R\sin (\phi -\psi ),  \nonumber \\
\phi ^{\prime } &=&-\cos (\phi -\psi ),  \label{AmplitudePhase} \\
\psi ^{\prime }+\Omega +R^2 &=&\cos (\phi -\psi ).  \nonumber
\end{eqnarray}
If we set $\Phi =\phi -\psi $, then we have
\begin{eqnarray}
\Phi ^{\prime }-\Omega -R^2 &=&-2\cos \Phi ,  \nonumber \\
R^{\prime } &=&-R\sin \Phi .  \label{AmplitudePhaseCompact}
\end{eqnarray}
Taking into account  equations~(\ref{AmplitudePhaseCompact}) we have
the conservation law
\begin{equation}
\cos \Phi =\frac{R^{2}}{4}+\frac{\Omega }{2}.  \label{SecondConsevationLaw}
\end{equation}
Substituting~(\ref{SecondConsevationLaw}) into the second equation
of~(\ref {AmplitudePhaseCompact}) and subsequent integration gives
the following expression for $R$:
\begin{equation}
R^2=\frac{2\beta^2}{\Omega +2\cosh \left\{ \beta (\eta -\eta
_{0})\right\} }, \label{r_solution}
\end{equation}
where $\beta =\sqrt{4-\Omega^2}$. The right-hand side is positive
real-valued for all $\eta $ if $-2<\Omega <2$. Using the
conservation law~(\ref{SecondConsevationLaw}) we obtain an
expression for $\Phi $:
\begin{equation}
\Phi =2\arctan \left( \frac{2-\Omega }{\beta }\tanh \left\{ \frac{1}{2}\beta
(\eta -\eta _{0})\right\} \right) .  \label{Phi_expression}
\end{equation}
Now we integrate $\phi ^{\prime }=-\cos \Phi $ and find
\[
\phi =-\frac{\Omega }{2}(\eta -\eta _{0})-\arctan \left(
\frac{2-\Omega }{ \beta }\tanh \left\{ \frac{\beta }{2}(\eta -\eta
_{0})\right\} \right) .
\]
Finally, we determine $\psi $:
\begin{eqnarray*}
\psi  &=&\phi -\Phi  \\
&=&-\frac{\Omega }{2}(\eta -\eta _{0})-3\arctan \left(
\frac{2-\Omega }{ \beta }\tanh \left\{ \frac{\beta }{2}(\eta -\eta
_{0})\right\} \right) .
\end{eqnarray*}
This pulse exists only if value of the parameter $\Omega $ is inside
the interval $-2<\Omega <2$. The maximal value of the amplitude of
this solitary solution is
\[
A=\sqrt{2(2-\Omega )}.
\]
The phases $\phi $ and $\psi $ are nonlinear. Their behavior is
asymptotically linear as $\eta \rightarrow \pm \infty $. If $\Omega =0$,
then the limiting values of the phases satisfy
\begin{eqnarray*}
|\phi (\infty )-\phi (-\infty )| &=&\pi /2, \\
|\psi (\infty )-\psi (-\infty )| &=&3\pi /2.
\end{eqnarray*}

The total energy of the solitary wave is distributed among the
counter-propagating fields and the medium polarization. Here we
study energy partition among these components. Using equations
(\ref{26.3}), (\ref{26.3.1}) and conditions as
$|\eta|\rightarrow\infty$, one can show that
\begin{equation}
f_a(\eta) = -\frac{1}{v} f_s(\eta) = -\frac{f_0}{v}F_{\Omega}(\eta).
\end{equation}
We are interested in the energies of the dimensionless fields $f_1$, $f_2$,
and of the polarization $q$.
\begin{eqnarray*}
f_1 & = & \frac{1}{2}(f_a - f_s) = -\frac{f_0}{2}\left(\frac{1}{v} +
1\right)F_{\Omega} \\
f_2 & = & -\frac{1}{2}(f_s + f_a) = \frac{f_0}{2}\left(\frac{1}{v} -
1\right)F_{\Omega} \\
q & = & q_0 Q
\end{eqnarray*}
Finally, for electric fields $e_1,e_2$ and for polarization $p$ we have
\begin{eqnarray}
e_1 &=& -\frac{1}{2\sqrt{\mu}}\left( \frac{2
v^2}{1-v^2}\right)^{3/4} \left( \frac{1}{v} + 1\right) \frac{\sqrt2~
\beta}{\sqrt{\Omega  + 2 \cosh\left\{\beta(\eta -
\eta_0)\right\}}}~e^{i\alpha_{e}}, \label{e1.final}
\\
e_2 &=& \frac{1}{2\sqrt{\mu}}\left( \frac{2 v^2}{1-v^2}\right)^{3/4}
\left( \frac{1}{v} - 1\right) \frac{\sqrt2~ \beta}{\sqrt{\Omega +2
\cosh\left\{\beta(\eta - \eta_0)\right\}}}~e^{i\alpha_{e}};
\label{e2.final}
\\
p &=& \frac{1}{\sqrt{\mu}}\left( \frac{2 v^2}{1-v^2}\right)^{1/4}
\frac{\sqrt2 ~\beta}{\sqrt{\Omega + 2 \cosh\left\{\beta(\eta -
\eta_0)\right\}}}~e^{i\alpha_{p}},  \nonumber \\
\alpha_{e} &=& \delta\tau -\frac{\Omega}{2}(\eta - \eta_0) -
\arctan\left(
\frac{2-\Omega}{\beta}\tanh\left\{\frac{\beta}{2}(\eta-
\eta_0)\right\}\right)  \nonumber \\
\alpha_{p} &=& \delta\tau -\frac{\Omega}{2}(\eta - \eta_0) -
3\arctan\left(\frac{2-\Omega}{\beta}\tanh\left\{\frac{\beta}{2}(\eta
- \eta_0)\right\}\right).  \label{q.final}
\end{eqnarray}

The energy of a solitary wave is
\begin{equation}
E_R = \int\limits_{-\infty}^{+\infty}|F_{\Omega}(\eta)|^2 d\eta =
8\arctan \sqrt{\frac{2-\Omega}{2+\Omega}}.
\end{equation}
Finally we have the energies of electric fields and polarization
\begin{eqnarray}
E_{e_1} &=& \frac{f_0^2}{4}\left(\frac{1}{v} + 1\right)^2 E_R, \nonumber\\
E_{e_2} &=& \frac{f_0^2}{4}\left(\frac{1}{v} - 1\right)^2 E_R, \label{energies}\\
E_p &=& q_0^2 E_R.\nonumber
\end{eqnarray}
Ratios of energies in different fields as well as polarization have the
following form
\begin{eqnarray}
\frac{E_{e_1}}{E_{e_2}} &=& \left(\frac{1+v}{1-v}\right)^2, \\
\frac{E_{e_1}}{E_p} &=& \frac{1}{2}\frac{(1+v)}{(1-v)}, \\
\frac{E_{e_2}}{E_p} &=& \frac{1}{2}\frac{(1-v)}{(1+v)}.
\end{eqnarray}
Therefore, energy partitioning is determined solely by the parameter $v$,
which is a dimensionless combination of main system parameters.

\section{Numerical simulation}

The shape and phase of the incident optical pulse are controllable in a real
experimental situation. To model pulse dynamics in the Bragg grating it is
natural to consider asymptotic mixed initial-boundary value problem for
equations~(\ref{SVEPEquations}). We define initial conditions as
\begin{equation}
p(\zeta,\tau)\rightarrow 0,~~~e_1(\zeta,\tau)\rightarrow
0,~~~e_2(\zeta,\tau)\rightarrow 0,~~~\tau \rightarrow -\infty,
\end{equation}
with no incident field at the right edge of the sample, and with incident
field at the left edge defined as follows:
\begin{eqnarray}
&e_1& (-10,\tau) = w\exp(i\theta),  \nonumber \\
&w& = 4.0\exp\left[-\frac{1}{2}\left(\frac{\tau
-3.0}{1.5}\right)^2\right] ,\;\; \theta =
\arctan\left(\tanh\left[1.5\left(\tau - 3.0\right)\right] \right).
\end{eqnarray}
In our case the spatial simulation domain was chosen as $[-10,40]$.
Parameters $\Delta -\delta$ and $\mu$ were given values
\begin{equation}
\Delta = 0,\;\; \delta = 1,\;\; \mu = 1.
\end{equation}

As one can see, we gave the initial pulse the same configuration of
phase in a topological sense as would be found in a solitary wave
solution. This point is important, because otherwise the phase
difference cannot relax to the symmetry of the stationary wave which
is revealed in~(\ref{Phi_expression}). As a result, in numerical
simulations it is difficult to achieve solitary wave type dynamics
of the pulse without the proper phase configuration of the incident
pulse.

The second field boundary condition was set to be
\begin{equation}
e_2 (40,\tau) = 0.
\end{equation}

\begin{figure}[hbt]
\centering
\includegraphics[width=4.0in]{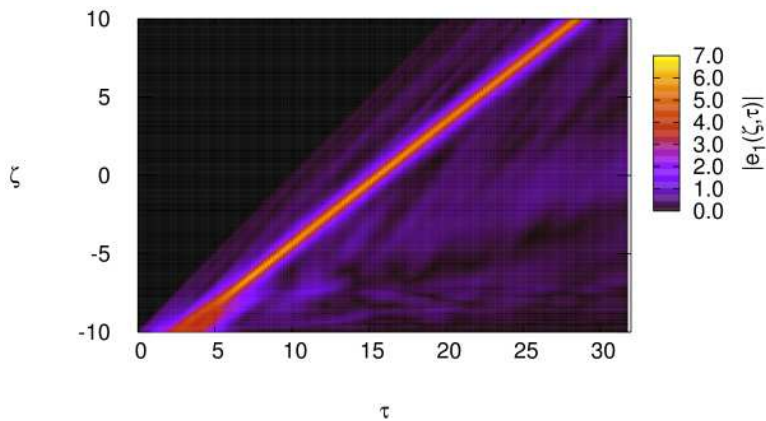}
\caption{Propagation of pulse. The first experiment. Mapping of the
$|e_1 ( \protect\zeta,\protect\tau)|$ surface.} \label{f1.1.map}
\end{figure}

\begin{figure}[hbt]
\centering
\includegraphics[width=4.0in]{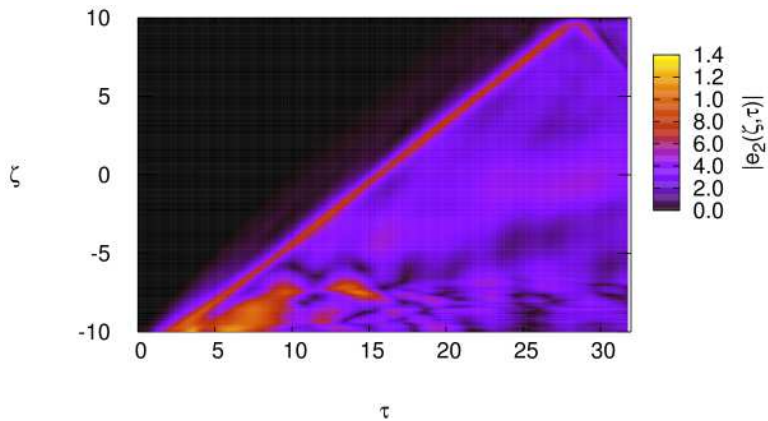}
\caption{Propagation of pulse. The first experiment. Mapping of the
$|e_2 ( \protect\zeta,\protect\tau)|$ surface.} \label{f2.1.map}
\end{figure}

\begin{figure}[hbt]
\centering
\includegraphics[width=4.0in]{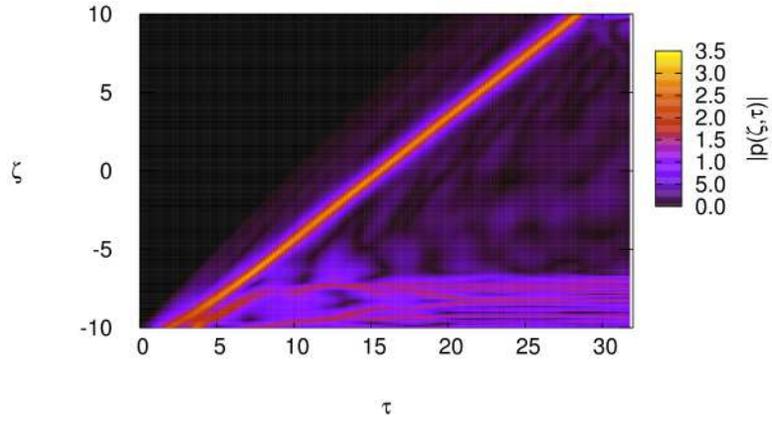}
\caption{Propagation of pulse. The first experiment. Mapping of the
$|p( \protect\zeta,\protect\tau)|$ surface.} \label{q.1.map}
\end{figure}

In the first experiment we injected a pulse, which is relatively
close to the solitary wave solution. The results are shown in
Figs.~\ref{f1.1.map}-\ref{q.1.map}, which clearly represent two
stages of the pulse evolution.
In the first stage ($t \le 7$) of evolution we observed fast excess
energy damping in radiation of quasi-linear waves in both directions
and relaxation to a solution roughly similar to the stationary
solution. Then we observed a stage of pulse shape refinement ($7 < t
< 30$) with consequent propagation of the solution very close to
(\ref {r_solution}). The lower part of the figure~\ref{q.1.map}
shows the creation of multiple spatially frozen ``hot spots'' of the
medium polarization. In these hot spots the energy of oscillations
is arrested in the grating. This ``stopping light'' phenomenon is
due to self-modulation of the medium polarization caused by
nonlinear effects which we shall discuss in more detail elsewhere.

\begin{figure}[hbt]
\centering
\includegraphics[width=4.0in]{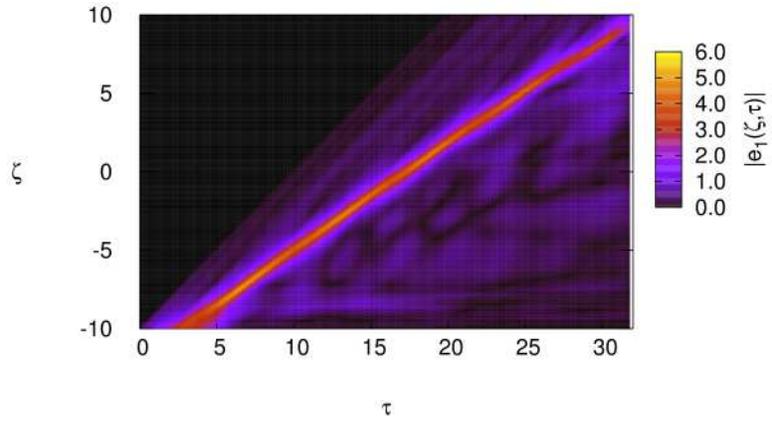}
\caption{Propagation of pulse. The second experiment. Mapping of the
$|e_1 ( \protect\zeta,\protect\tau)|$ surface.} \label{f1.2.map}
\end{figure}

\begin{figure}[hbt]
\centering
\includegraphics[width=4.0in]{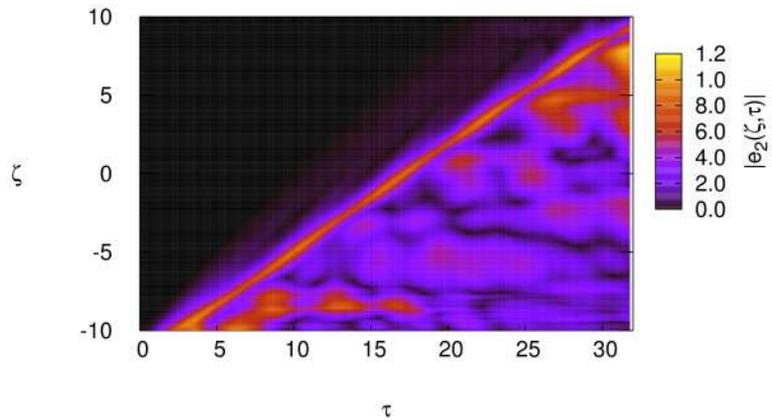}
\caption{Propagation of pulse. The second experiment. Mapping of the
$|e_2 ( \protect\zeta,\protect\tau)|$ surface.} \label{f2.2.map}
\end{figure}

\begin{figure}[hbt]
\centering
\includegraphics[width=4.0in]{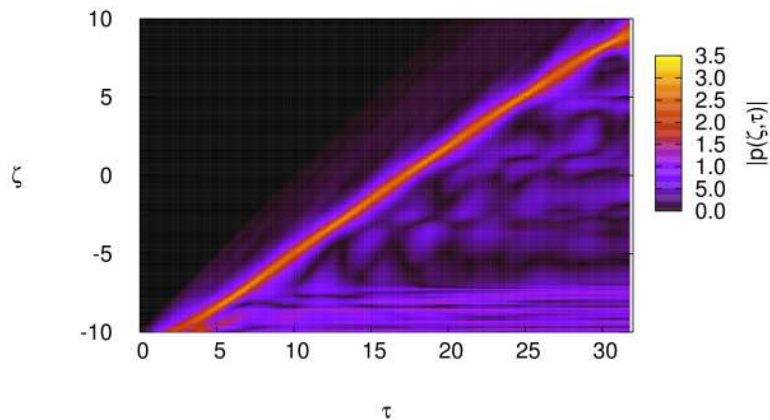}
\caption{Propagation of pulse. The second experiment. Mapping of the
$|p( \protect\zeta,\protect\tau)|$ surface.} \label{q.2.map}
\end{figure}

During the second experiment we used a pulse of lower amplitude:
\begin{eqnarray}
e_1(-10,\tau) & = & w\exp(i\theta),  \nonumber \\
w & = & 3.3\exp\left[-\frac{1}{2}\left(\frac{\tau
-3.0}{1.5}\right)^2\right],
\nonumber \\
\theta & = & \arctan\left(\tanh\left[1.5\left(\tau -
3.0\right)\right] \right).
\end{eqnarray}
Results are represented in Figs.~\ref{f1.2.map}-\ref{q.2.map}.
Deviation of the incident pulse from exact solitary wave solution
was stronger in the second numerical experiment. In this case,
computer simulations clearly demonstrate oscillatory behavior of the
pulse dynamics. Study of the origin of such oscillations will also
be presented in a separate publication. In both cases we observed a
phenomenon similar to self-induced transparency in nanocomposite
materials described in~\cite{R16} . These solitary wave solutions
represent the phenomenon of nonlinear light trapping when the wave
of medium polarization is bound by two optical light fields
propagating in the same direction. In the linear limit these two
optical fields are propagating in opposite directions.

\section{Conclusion}

We studied, both analytically and numerically, solitary waves
describing coupled medium polarization and light propagation in
Bragg gratings with thin films containing metallic nanoparticles. We
observed self-induced transparency-like phenomena and formation of
polarization ``hot spots''. The formation of solitary wave solutions
occurs when energy of an incident pulse exceeds some threshold
value. The lower boundary value  of this threshold can be evaluated
as sum of energies~(\ref{energies}) of electric and plasmonic
components of coupled wave.

Stability of  nonlinear solitary waves is an important problem which
can be addressed by study of collision properties  and analysis of
localized modes of such waves. This problem is the subject of our
future investigations.

\begin{acknowledgments}
We would like to thank B.I.~Mantsyzov, A.A.~Zabolotskii, J-G.~Caputo,
M.G.~Stepanov and R.~Indik for enlightening discussions. AIM and KAO are
grateful to the Laboratoire de Math\'{e}matiques, INSA de Rouen and the
University of Arizona for hospitality and support. This work was partially
supported by NSF (grant DMS-0509589), ARO-MURI award 50342-PH-MUR and State
of Arizona (Proposition 301), RFBR grants 06-02-16406 and 06-01-00665-a,
INTAS grant 00-292, the Programme ``Nonlinear dynamics and solitons'' from
the RAS Presidium and ``Leading Scientific Schools of Russia'' grant. KAO
was supported by Russian President grant for young scientists MK-1055.2005.2.
\end{acknowledgments}

\end{document}